\documentclass[conference]{IEEEtran}

\usepackage{float}
\usepackage{caption}
\usepackage{trivfloat}
\usepackage{amsmath,amssymb,amsfonts}
\usepackage{cite}
\usepackage{algpseudocode}
\usepackage{graphicx}
\graphicspath{{./Images/}}
\usepackage{textcomp}
\usepackage{tabularx}
\usepackage[table,xcdraw,dvipsnames]{xcolor}
\usepackage{booktabs}
\usepackage{listings}
\usepackage{multirow}
\usepackage{colortbl}
\usepackage{amsthm}
\newtheorem{definition}{Definition}

\usepackage{mdframed}
\usepackage{soul}
\usepackage{comment}

\def\BibTeX{{\rm B\kern-.05em{\sc i\kern-.025em b}\kern-.08em
    T\kern-.1667em\lower.7ex\hbox{E}\kern-.125emX}}
    
\trivfloat{Listing}

\newcommand{\originalListing}[0]{
  \begin{Listing}[b!]
    {\vskip -0.5em}
    \centering
    \lstinputlisting[language=Java]{code/original.java}
    \caption{Original code for the motivating example}
    \label{lst:OrigCode}
    {\vskip -0.5em}
  \end{Listing}
}

\newcommand{\predicateListing}[0]{
  \begin{Listing}[b!]
    {\vskip -0.5em}
    \centering
    \lstinputlisting[language=Java]{code/predicate.java}
    \caption{Predicates in Listing~\ref{lst:OrigCode} extracted to boolean variables.}
    \label{lst:Pred_Code}
    {\vskip -0.5em}
  \end{Listing}
}

\newcommand{\instrumentedListing}[0]{
  \begin{Listing}[b!]
    {\vskip -0.5em}
    \centering
    \lstinputlisting[language=Java]{code/instrumented.java}
    \caption{Instrumented version of the code that belongs to the \texttt{format} method from Listing~\ref{lst:OrigCode}
    }
    \label{lst:GSA_Code}
    {\vskip -0.5em}
  \end{Listing}
}

\begin{document}
\title{Improving Fault Localization by Integrating Value and Predicate Based Causal Inference Techniques}

 \author{
   \IEEEauthorblockN{
     Yi\u{g}it K\"{u}\c{c}\"{u}k,
   }
   \IEEEauthorblockA{
     \emph{Department of Computer and Data Sciences} \\
    \emph{Case Western Reserve University}\\
    Cleveland, OH, USA \\
    yxk368@case.edu
    }
    \and
     \IEEEauthorblockN{Tim A. D. Henderson,}
    \IEEEauthorblockA{
      Google Inc. \\
     Mountain View, CA, USA \\
      tadh@google.com
    }
    \and
    \IEEEauthorblockN{Andy Podgurski}
    \IEEEauthorblockA{
     \emph{Department of Computer and Data Sciences } \\
    \emph{Case Western Reserve University}\\
    Cleveland, OH, USA \\
   podgurski@case.edu
    }
}

\maketitle

\begin{abstract}

  Statistical fault localization (SFL) techniques use execution profiles and success/failure information from software executions, in conjunction with statistical inference, to automatically score program elements based on how likely they are to be faulty.  SFL techniques typically employ one type of profile data: either coverage data, predicate outcomes, or variable values.  Most SFL techniques actually measure correlation, not causation, between profile values and success/failure, and so they are subject to confounding bias that distorts the scores they produce.  This paper presents a new SFL technique, named \emph{UniVal}, that uses causal inference techniques and machine learning to integrate information about both predicate outcomes and variable values to more accurately estimate the true failure-causing effect of program statements.  \emph{UniVal} was empirically compared to several coverage-based, predicate-based, and value-based SFL techniques on 800 program versions with real faults.

\end{abstract}


\section{Introduction}

There has been a vast amount of research on \emph{automated software fault localization} (AFL) \cite{jones2002visualization,liblit2005scalable, zeller2002simplifying, wong2016survey}, which seeks to automate all or part of the process of locating the software faults that are responsible for observed software failures.  \emph{Statistical fault localization} (SFL) or \emph{spectrum-based fault localization} \cite{jones2002visualization, liblit2005scalable, abreu2009practical, wong2016survey} comprises a large body of AFL techniques that apply statistical measures of association --- some computed with machine learning or data mining techniques --- to execution data (execution profiles or \emph{spectra}) and to failure data (e.g., pass/fail labels) in order to compute putative measures, called \emph{suspiciouness scores}, of the likelihood that individual program statements or other program elements are responsible for observed failures. These scores are used to guide developers to faults, e.g., by using them to highlight suspicious statements in graphical displays of code \cite{jones2002visualization} or to rank statements for inspection \cite{wong2016survey}.  Statistical fault localization is also the first step in a number of automated program repair techniques \cite{nguyen2013semfix, assiri2017fault}.

SFL techniques are applicable when the data they require are readily and cheaply available in sufficient quantity.  In particular, they generally require data from a diverse and penetrating set of tests or operational executions that includes significant numbers of both failures and successes.  Such data might be available, for instance, from deployed software that is instrumented to record profile data and that is equipped with a mechanism by which users may report failures they encounter.

It seems fair to say that statistical fault localization research stands at a crossroads.  Although a large number of SFL techniques have been proposed, to our knowledge none of them consistently locates faults with enough precision to justify its widespread use in industry.  In part, this is because most such techniques rely on just one source of information about internal program behavior: code coverage profiles, or, similarly, recorded outcomes of program predicates such as those that control the execution of conditional branches and loops.  Recent work has sought to overcome this limitation, e.g., by employing \emph{value profiles/spectra} (profiles of variable values) \cite{xie2005checking, jeffrey2008fault,gore2011statistical} or by combining different kinds of information pertinent to fault localization, such as coverage-based SFL scores, textual similarity measures, and fault-proneness predictions based on static program analysis \cite{li2019deepfl,xuan2014learning,sohn2017fluccs,zou2019empirical}.  Another problem with existing SFL techniques is that many of them suffer from \emph{confounding bias}, which can cause a correct statement to appear suspicious because of another, faulty statement that influences its execution.  Recent work has also sought to overcome this problem by employing \emph{causal inference} techniques \cite{baah2010causal, baah2011mitigating, gore2012reducing,shu2013mfl,bai2017causal}.

Existing SFL techniques are based on analysis of code coverage profiles or predicate profiles, on one hand, or of value profiles, on the other hand.  To our knowledge, no previous technique is both coverage-based (or predicate-based) and value-based.  While two techniques, one of each kind, can be combined simply by taking the average or maximum of the scores they produce for each potential fault location, this will not in itself properly address confounding bias.

This paper proposes a new approach to statistical fault localization that integrates information about both predicate outcomes and variable values, and that does so in a principled way that controls for confounding bias.  The key to this approach, which we call \emph{UniVal}, is to transform the program under analysis so that branch and loop predicate outcomes become variable values, so that one causally sound value-based SFL technique can be applied to both variable assignments and predicates.  This paper reports on an large-scale empirical evaluation of \emph{UniVal} involving the latest version (2.0.0) of the widely used Defects4J evaluation framework \cite{just2014defects4j}, which contains seventeen software projects and 835 program versions containing actual faults. In this study, \emph{UniVal} was compared to several coverage-based, predicate-based, and value-based SFL techniques.  \emph{UniVal} substantially out-performed each of these techniques.  To the best our knowledge this is the only fault localization study that has used all the programs in the latest version of Defects4J, and it uses the largest number of real programs and faults of any study.   We also report empirical results characterizing the relationship between the cost of fault localization and an important property of data in a causal inference study called \emph{covariate balance}.  These results help explain the effectiveness of \emph{UniVal}.

\originalListing
\predicateListing
\instrumentedListing

\section{Motivating Example} 
\label{MotEx}

To illustrate our technique, we now apply \emph{UniVal} to locate a real software fault. This example is from the 62nd faulty version of Google's Closure Compiler (\texttt{Closure 62}) from the widely used Defects4J\cite{just2014defects4j} repository (v2.0). This fault exists in the \texttt{format} method, which is located between lines 66-111 in the \texttt{LightweightMessageFormatter} class. The faulty segment of the code is depicted in Listing~\ref{lst:OrigCode} with some elements omitted for brevity, and with the lines of code renumbered.

The method \texttt{format} is intended to compose an error message for a script written in Javascript. As input parameters, \texttt{format} takes a \texttt{JSError} object named \texttt{error}, which contains information about the nature of the error, and a boolean value named \texttt{warning}, which indicates whether the error message should be formatted as a warning message. However, this method has a faulty predicate at line 5 (``$<$'' is used instead of ``$\leq$'') that prevents the error column number, \texttt{charno}, from being displayed correctly for some inputs to the method. The fix for the faulty predicate expression is shown in a comment on line 4.

The code in Listing 1 illustrates how confounding bias  \cite{pearl2009causality} may arise in fault localization: the outcome of the faulty predicate at line 5 confounds the causal effects on program failure of the statements at lines 6-14. This implies that unless we adjust for confounding, the suspiciousness scores calculated for the variables and predicates located at those lines are likely to be biased and distorted.

In order to apply \emph{UniVal} and other fault localization techniques to the code in Listing 1, we first used a tool we developed named \emph{PredicateTransformer} to transform the predicates in branch and loop conditions into assignment statements that assign the results of evaluating the predicates to new boolean variables. Note that each predicate in a compound boolean expression is transformed in this way. The transformed code is depicted in Listing~\ref{lst:Pred_Code}. 

Next, we used our instrumentation tool \emph{GSA\_Gen} on the predicate-transformed program to generate our own variant of the \emph{gated static single-assignment (GSA) form}  \cite{Ottenstein1990} representation of the program. This tool collects information about data dependencies between variables, and it also instruments the program to record the runtime values of assignment targets in a dictionary data structure that is maintained by a Java class we created named \emph{CollectOut}.  The variable values are recorded by inserting calls to a static method \emph{CollectOut.record()} (referred to as \texttt{record()} in Listing~\ref{lst:GSA_Code}), which takes the following parameters in the given order: package name, class name, method name, line number encountered in the original code, code block number, name in GSA form, value to be recorded, and the version of the program variable.  The GSA version of the code in Listing 1 is shown in Listing~\ref{lst:GSA_Code}.

Consider what happens when \emph{GSA\_Gen} encounters the assignment statement \texttt{P3\_1 = (0 <= charno)} at line 8 of Listing~\ref{lst:Pred_Code}.  The tool then: inserts the proper GSA variable version increments; inserts a call to \emph{CollectOut.record()} to record the value of \texttt{P3\_1} and the other parameters mentioned above; and adds a key-value pair \texttt{[P3\_1\_i,charno\_j]} to the dictionary, where \texttt{i} and \texttt{j} are the respective variable versions in the assignment statement. The value(s) for key \texttt{P3\_1\_i} are later used for confounding adjustment (see Section~\ref{Method}).

We tested the GSA transformed program with the developer-written tests in the Defects4J suite, using the \emph{test} command to run these tests. Out of the 6050 tests, two failed with assertion errors. We applied two of the coverage-based statistical fault localization metrics that performed the best in recent studies\cite{abreu2009new,pearson2017evaluating,wong2014dstar,lucia2014extended}, namely, Ochiai \cite{abreu2009practical} and DStar\cite{wong2014dstar}, to the code coverage and failure data for all of the tests.  We then ranked the variables and predicates of the faulty class, \texttt{LightweightMessageFormatter}, in nonincreasing order of their suspiciousness scores (with ties receiving the average rank for all tied variables). The Ochiai and DStar metrics produced \emph{identical} rankings. The faulty predicate \texttt{P3\_0} at line 13 in Listing~\ref{lst:GSA_Code} (line 5 in Listing~\ref{lst:OrigCode}) received the same score as 23 other variables and ranked in 17th place in the resulting suspiciousness list, which included all of the variables and predicates within the \textbf{if} branch in the faulty method. The low rank for the faulty predicate is likely due to confounding, which the Ochiai and DStar metrics don't adjust for.

We next applied \emph{UniVal} to the same transformed program. We input the data recorded by our run-time library \emph{CollectOut} together with the values of a binary outcome variable $Y$, which indicates whether the program executions failed ($1$) or succeeded ($0$), to our method for calculating suspiciousness scores. This code executes the \emph{analysis phase} of \emph{UniVal}, which is described in Section~\ref{Method}. Finally, we mapped these scored statements back to the original program.

The suspiciousness list produced with \emph{UniVal} contained unique scores for each variable and predicate.  The faulty predicate \texttt{P3\_0}, at line 13 in Listing~\ref{lst:GSA_Code} (line 5 in Listing~\ref{lst:OrigCode}), was ranked highest among all the predicates with a score of 0.086, and it ranked 5th among all assignment targets.  The variables \texttt{charno\_2} and \texttt{charno\_3} recorded at lines 14 and 15 of Listing~\ref{lst:GSA_Code}, which are recorded for possible use in simulating a GSA-form $\phi_{\mathit{if}}$ function (see Section~\ref{GSA}), represent the value of the variable \texttt{charno} in lines 12 and 13. They would both be ranked 1st overall, with a score of 0.49, if we included them in the ranking.\footnote{In our empirical evaluation, we assign scores to only the predicates in branch conditions.} The non-faulty predicate \texttt{P4\_0}, at line 21 in Listing~\ref{lst:GSA_Code} (line 6 in Listing~\ref{lst:OrigCode}), was ranked 18th with a score of 0.001.

We performed an extra step to the experiment to illustrate that the adjustment for confounding in \emph{UniVal} indeed makes a difference. We replicated the steps described previously for using \emph{UniVal} with a minor change for the correct predicate \texttt{P4\_0}, which is located at line 21 in Listing~\ref{lst:GSA_Code}. We removed \texttt{P3\_0} (at line 13 in Listing~\ref{lst:GSA_Code}) from the set of adjustment variables (covariates) in the random forest model for \texttt{P4\_0}.
Consequently, the score for  \texttt{P4\_0} increased to 0.01, which changed its rank to 14th. The new score and rank of \texttt{P4\_0} were higher than those of the faulty predicate \texttt{P3\_0}, whose score and rank declined to 0.005 and 16th.
Therefore, the suspiciousness scores for the variables were in fact distorted by the lack of adjustment for confounding bias.

\section{Background}
\label{background}

\subsection{Causal Inference}
\label{CausalInference}

In this section we provide background on statistical causal inference required to understand our technique, \emph{UniVal}. Statistical causal inference is concerned with estimating without bias the average causal effect of a treatment variable upon an outcome variable.  A \emph{treatment variable} or \emph{exposure variable} $T$ is a variable that an investigator could, at least in principle, intervene upon to change its value.  For example, in SFL $T$ might represent the outcome of a branch predicate, which a developer could change in a debugger.  An \emph{outcome variable} $Y$ is an observable variable of interest to an investigator, such as an indicator of whether a program execution fails or not.  In statistical causal inference, one is concerned with outcomes for individuals or units in a population, and it is often convenient to denote the outcome for a unit $i$ by $Y_i$.  In SFL, the units are typically program executions.

A key concept in causal inference is that of a \emph{counterfactual outcome}, which is a value that the outcome variable $Y$ could take on if, counter to the facts, the treatment variable $T$ took on a value $t'$ different from the value $t$ that it actually took on.  A very similar concept is that of a \emph{potential outcome}, which is a value that the outcome variable could take on if the treatment variable were assigned a particular value.  As is common in the causal inference literature, we shall use the terms ``counterfactual outcome'' and ``potential outcome'' interchangeably.  In the influential Neyman-Rubin causal modeling framework \cite{imbens2015causal}, a distinct \emph{potential outcome random variable} $Y^{T=t}$ exists for each possible value $t$ of the outcome variable.

In principle, for each unit $i$ and each pair of treatment values $t$ and $t'$ there is an \emph{individual causal effect} $\delta_i = y^t_i - y^{t'}_i$, where $y^t_i$ and $y^{t'}_i$ are values of the potential outcome random variables $Y^{T=t}_i$ and $Y^{T=t'}_i$, respectively.  Conceptually, the \emph{average causal effect} (ACE) of $T$ on $Y$ over a population is the expected value $\mathrm{E}[\delta]$ of the individual causal effects $\delta_i$ over that population.  However, in general it is not possible to compute the $\delta_i$, because at most one of $Y^{T=t}_i$ and $Y^{T=t'}_i$ is observed for each $i$, namely, the potential outcome under the actual treatment.  This is known as the \emph{Fundamental Problem of Causal Inference} \cite{holland1986statistics}. In this sense, causal inference is a \emph{missing data problem}.  In statistical causal inference, this problem is circumvented by estimating average causal effects from a study sample of units and by ``borrowing information'' from each unit in the sample.

In a randomized experiment, which is an interventional study in which the the investigator assigns treatments randomly to units, the missing potential outcomes are missing completely at random, and the average causal effect $\mathrm{E}[\delta] = \mathrm{E}[Y^{T=t} - Y^{T=t'}] = \mathrm{E}[Y^{T=t}] - \mathrm{E}[Y^{T=t'}]$ can be estimated by computing the average outcomes in the treatment groups with $T=t$ and $T=t'$ and taking their difference \cite{hernan2018causal}.  However, in observational studies generally and in SFL applications in particular, treatment assignment is usually \emph{not} randomized, and hence the difference of the treatment group averages is often a biased estimate of the average causal effect. In SFL, for example, whether a particular statement is executed during a program run or whether a given variable is assigned a specific value depends on the behavior of other statements, which may also affect whether the program fails.

Different types of bias can affect a causal effect estimate \cite{hernan2018causal}.  The best known form of systematic bias and the one that will be addressed in this paper is \emph{confounding bias} (or simply \emph{confounding}). Confounding is bias due to the presence of a variable that is a \emph{common cause} of the treatment variable and the outcome variable.  Confounding bias is best explained in terms of a causal graph, and such graphs are also used to identify \emph{confounders}, which are variables that can be statistically adjusted for during causal effect estimation in order to reduce or eliminate confounding bias.  A \emph{causal directed acyclic graph} or \emph{causal DAG} is a DAG in which the vertices represent causal variables and in which there is a directed edge $(A,B)$ or $A  \rightarrow B$ between two variables $A$ and $B$ only if $A$ is known or assumed to be a cause of $B$.

\begin{figure}
\centering
  \includegraphics[width=0.45\columnwidth]{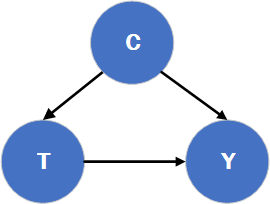}
  \caption{Example Confounding DAG}
  \label{Confounding_DAG}
  {\vskip -2em}
\end{figure}

Figure~\ref{Confounding_DAG} is a very simple example of a DAG involving three variables, $T$, $C$, and $Y$.  The edge $T \rightarrow Y$ represents the direct causal effect of $T$ on $Y$.  The DAG indicates that $C$ confounds this effect, because $C$ is a common cause of $T$ and $Y$.  The noncausal path $T \leftarrow C \rightarrow Y$, which is called a \emph{backdoor path} because it begins with an arrow into the treatment variable, represents a biasing flow of statistical association from $T$ to $Y$ via $C$.  As a result of this flow, a naive, unadjusted estimate of the ACE would mix the causal effect of $T$ on $Y$ with the association ``carried'' by the backdoor path.  This implies that to obtain an unbiased estimate of the ACE, $C$ must be adjusted for.

One way to obtain an unbiased estimate of the ACE is to \emph{block} ($d$-separate \cite{pearl2009causality}) the backdoor path $T \leftarrow C \rightarrow Y$ by conditioning on the value of $C$ during the analysis, e.g., by computing causal effect estimates separately for each level or stratum of $C$ and then combining them via a weighted average to obtain an estimate of the population ACE.  Causal inference theory provides a number of results that characterize the sets of variables that may be used for confounding adjustment, in terms of the structure of a causal DAG.  For example, the \emph{Backdoor Adjustment Theorem} \cite{pearl2009causality} states that a set $Z$ of variables that blocks every backdoor path between the treatment variable and the outcome variable is sufficient for confounding adjustment.

Another way to estimate the average causal effect, which we employ in this paper, is to interpolate or predict the missing counterfactual outcome $Y^{T=t}_i$ or $Y^{T=t'}_i$ for each unit based on the data for all the units in the study sample and then to plug in the predicted value in the formula for the individual causal effect $\delta_i$.

\begin{figure*}
\centering
  \fbox{\includegraphics[width=1.9\columnwidth]{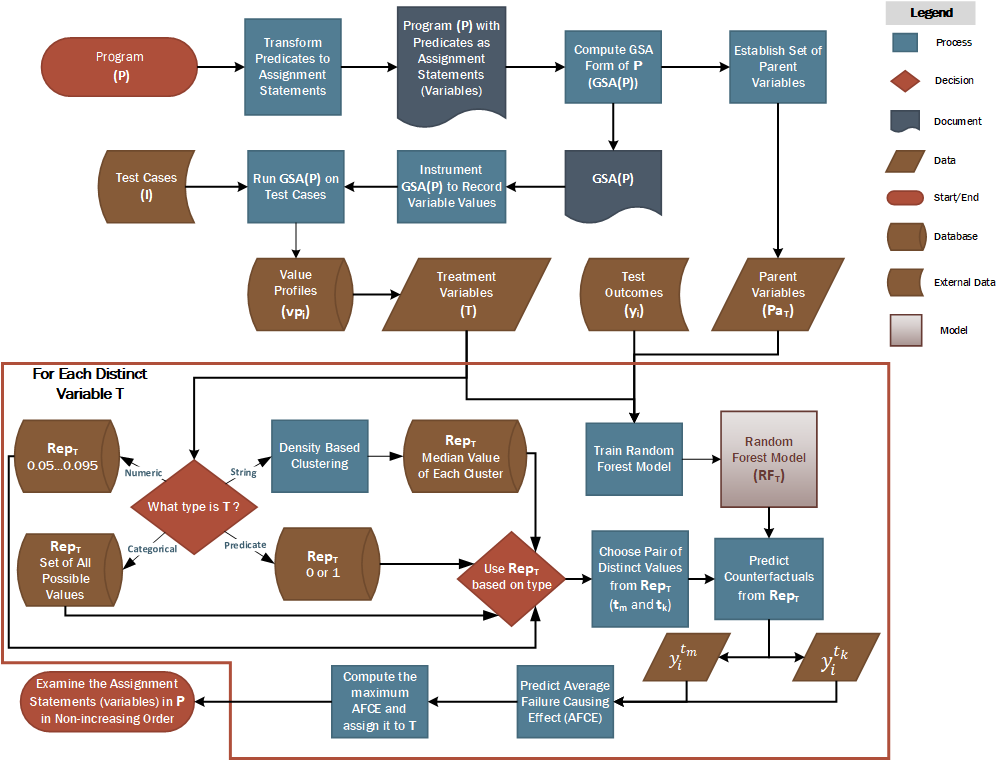}}
  \caption{Causal fault localization method \emph{UniVal}}
  \label{MethodFig}
  {\vskip -1em}
\end{figure*}

\subsection{Gated Static Single Assignment Form}
\label{GSA}

\emph{Gated single assignment form} (GSA form) \cite{Ottenstein1990} is an extension of \emph{static single  assignment form} (SSA form) \cite{Cytron1991}.  SSA form is a specialized intermediate program representation that makes the data flow of a program explicit by ensuring that each variable is defined in exactly one location (hence the name static \emph{single} assignment form). When two or more definitions for a variable reach a single use a new definition is created to merge the reaching definitions by using a special ``pseudo-function'' called $\phi$ which ``picks'' the correct definition to use at runtime.

GSA replaces the $\phi$ function of SSA form with three gating functions $\phi_{\mathit{if}}$, $\phi_{\mathit{entry}}$, and $\phi_{\mathit{exit}}$ (alternatively, $\gamma$, $\mu$, and $\eta$, respectively).  $\phi_{\mathit{if}}$ represents the merging of control flow after an if-statement and takes as an argument the \emph{predicate} in the controlling if-statement, allowing $\phi_{\mathit{if}}$ to choose the correct value. $\phi_{\mathit{entry}}$ is similar to $\phi_{\mathit{if}}$ except it merges loop-carried variables at the top of iterative control structures.  Finally, $\phi_{\mathit{exit}}$ merges variables that are both \emph{live} at the exit of a loop and are \emph{modified} by that loop. Taken together these three new gating functions effectively embed the control dependence graph \cite{ferrante1987program} into the intermediate representation by linking the choice of the variable definition to use to the predicate which controls the computation.

The instrumentation used by  \emph{UniVal} is inspired by (and its placement is guided by) GSA form. If GSA form was directly converted into program instrumentation it would be expensive due to the extra runtime overhead implied by the gating functions $\phi_{\mathit{if}}$, $\phi_{\mathit{entry}}$, and $\phi_{\mathit{exit}}$ for choosing the right definitions.  Instead instrumentation is inserted to record values (and their controlling predicates) at the locations where the gating functions would have appeared in GSA form. (See Section~\ref{MotEx}.)  This permits us to determine causal parents needed to control for confounding in \emph{UniVal}.


\section{Method}
\label{Method}

In this section, we describe the \emph{UniVal} fault localization technique in detail.  \emph{UniVal} is based on predicting the values of individual counterfactual outcomes, using machine learning models that are trained on data from a sample consisting (ideally) of significant numbers of both passing and failing executions.  There is a separate model for each assignment to a program variable, including assignments inserted into branch predicates (e.g., lines 11-13 in Listing~\ref{lst:GSA_Code}).  The data required for each execution and each assignment includes the values of the treatment variable $T$, the outcome variable $Y$, and the set of covariates $X$. The treatment variable is the assignment target, the outcome variable indicates whether the execution passed or failed, and the covariates are the parents of the treatment variable (if any exist) in the causal DAG for the program being debugged.  For example, the recorded data for line 12 of Listing~\ref{lst:GSA_Code} includes the values of \texttt{P3\_1} and \texttt{charno} as well as the program outcome (pass or fail).  Note that each backdoor path (see Section~\ref{background}) in the causal DAG begins with an arrow $T\leftarrow P$, where $P$ is a parent of $T$.  Thus, by the Backdoor Adjustment Theorem (see Section~\ref{background}) the set $X$ of parents of $T$ blocks all backdoor paths from $T$ to $Y$, and therefore $X$ is sufficient for confounding adjustment.

Figure~\ref{MethodFig} depicts the operation of \emph{UniVal}. The first two steps in \emph{UniVal}, which together we call the \emph{instrumentation phase}, are source-to-source transformations of the program $\cal P$ to be debugged.  First, each predicate in a branch conditions of $\cal P$ is transformed into an assignment statement that assigns the value of the predicate to a new boolean variable.  (See for example lines 7-9 of Listing~\ref{lst:Pred_Code}, which correspond to line 5 of Listing~\ref{lst:OrigCode}.)  We created a prototype tool named the \emph{PredicateTransformer} for this task.  This tool also records information about each predicate, including the type of control statement it belongs to (e.g. while, for, if-else, if), the predicate expression, and the line number where it was encountered in the original Java file. Second, the resulting program is transformed by another prototype tool we created, named \emph{GSA\_Gen}, into the  specialized version of Gated Single Assignment form described in Section~\ref{GSA}. This entails inserting calls to a function that records the values of treatment variables and covariates as well as other information needed by our implementation.  At the same time, the causal parents of assignment targets (the variables whose values are used in the assignment) are determined.

The next phase of \emph{UniVal}, which we call the \emph{profiling phase}, involves executing the instrumented GSA version $GSA({\cal P})$ of the program on a set $\cal I$ of test cases or operational inputs in order to record the variable values and other information mentioned above.  Note that for a treatment variable $T$ assigned to in a loop, only the \emph{last} value assigned to $T$ is recorded, along with the corresponding covariate values.  It is assumed that the program outcomes (pass or fail) for these executions are already known or are determined prior to analysis.  

In \emph{UniVal}, assignment targets will have missing values in particular program runs (which we call NAs) if their assignment statements are not executed. If the treatment variable has no recorded values or just one unique value, it is omitted from \emph{UniVal}'s consideration.  Note that parts of a compound predicate expression might have missing values due to short-circuited evaluation.

The third phase of \emph{UniVal}, which we call the \emph{analysis phase}, involves fitting a counterfactual prediction model for each assignment target in $GSA({\cal P})$.  We adopted the counterfactual prediction approach to analysis described in \cite{podgurski2020counterfault}, although, unlike \emph{UniVal}, that work does not address predicates or strings.  In the current implementation of \emph{UniVal}, we employ random forest learners \cite{breiman2001random} as prediction models, because they are flexible enough to non-parametrically model a wide variety of relationships between the treatment variable and covariates, on one hand, and the outcome variable, on the other hand. Specifically, we used the \emph{Ranger} random forest package \cite{wright2015ranger}.  The model for a given variable assignment includes the treatment variable (the assigned variable) and the covariates (the used variables) as predictors, and the model is used to predict the counterfactual program outcomes (pass or fail) under different values of the treatment variable.

For each treatment variable $T$, a set $\mathit{Rep}_T$ of representative treatment values is chosen, and counterfactual outcomes are predicted for these treatments. $\mathit{Rep}_T$ is chosen differently based on the type of $T$.  For a boolean or categorical treatment variable $T$, $\mathit{Rep}_T$ contains all recorded values of $T$.  For a string variable, a clustering algorithm is first used to cluster the recorded treatment values, and then $\mathit{Rep}_T$ becomes the set of cluster IDs, which are treated like categorical values. We use the \emph{stringdist}\cite{van2019package} package to obtain a matrix of distances between values.  This matrix is input to the distance-based clustering algorithm  \emph{DBSCAN}\cite{hahsler2019dbscan} (with MinPts = $2*dim$, where $dim$ is the dimension of the dataframe).
For a numeric treatment variable $T$, $\mathit{Rep}_T$ consists of the $0.05, 0.15, \ldots, 0.95$ quantiles of the empirical distribution of the recorded values of $T$.  \emph{UniVal} does not currently handle other data types.

For each representative treatment value $t\in\mathit{Rep}_T$ and each complete input $i$ to $\cal P$, \emph{UniVal} predicts the counterfactual outcome $y^t_i$.  This is done by plugging $t$ into the model together with the covariate values $x_i$ recorded at the assignment to $T$ during execution of $\cal P$ on $i$. Note that \emph{UniVal} predicts counterfactual outcomes even for actual, recorded treatment-covariate combinations, that is, even when the true counterfactual outcome is known.
Given the set of predicted counterfactual outcomes for $T$, \emph{UniVal} computes for each $t\in\mathit{Rep}_T$ an estimate $\hat{\mathrm{E}}[Y^{T=t}]$ of the counterfactual mean $\mathrm{E}[Y^{T=t}]$, by averaging the predictions $\{ \hat{y}^t_i \}$.  The suspiciousness score for $T$ is set to the maximum, over all pairs $t,t'\in\mathit{Rep}_T$ of $\hat{\mathrm{E}}[Y^{T=t}]-\hat{\mathrm{E}}[Y^{T=t'}]$.  That is, the score is the maximum, over all pairs of representative treatment values for $T$, of the average failure-causing effect of assigning $t$ instead of $t'$ to $T$.

The final phase of \emph{UniVal}, which we call the \emph{localization phase}, involves employing the suspiciousness scores to assist developers in finding the cause or causes of failures observed when $\cal P$ was executed on the set of inputs $\cal I$.  Traditionally, this is done by ranking statements in non-increasing order of their suspiciousness scores and then having developers inspect statements in that order \cite{wong2016survey}.  Although this is convenient for evaluating SFL techniques --- and it is used for that purpose in this paper --- it has been argued that this is a simplistic approach to fault localization \cite{parnin2011automated,henderson2019evaluating}, which programmers are in many cases unlikely to follow (e.g., when many statements in a program get very high scores).  We envision \emph{UniVal} being used in combination with other sources of information, including developer knowledge and intuition, to effectively localize faults.

\section{Empirical Evaluation}
\label{Evaluation}

\subsection{Study Setup}
\label{StudSetup}

We empirically evaluated the fault localization performance of \emph{UniVal} in a substantial empirical study involving subject programs from the latest, expanded version (2.0.0) of the popular Defects4J evaluation framework \cite{just2014defects4j}. We compared the fault localization costs of \emph{UniVal} and several competing techniques: the non-interventional value-based techniques Elastic Predicates (ESP) \cite{gore2011statistical} and NUMFL (specifically the two variants NUMFL-DLRM and NUMFL-QRM) \cite{bai2017causal}; the non-interventional coverage-based technique of Baah \emph{et al.} \cite{baah2010causal}, which employs linear regression for causal inference; the interventional technique Predicate Switching \cite{zhang2006locating}, which alters conditional branching; and finally two well-known coverage-based SFL (CB-SFL) metrics that performed well in recent comparative studies \cite{abreu2009practical,pearson2017evaluating,lucia2014extended}, namely, Ochiai \cite{abreu2009practical} and D-Star (with star $=2$) \cite{wong2014dstar}. 

There is one important note concerning the implementation of Baah \emph{et al.}'s original technique based on linear regression \cite{baah2010causal}. With that technique, the only covariate in the regression model was a coverage indicator for the forward control dependence predecessor of the target statement.  For this study, we have modified Baah \emph{et al.}'s technique by including the variables used at the target statement as covariates.  We believe this is a notable improvement on the original technique. The modified version almost always performs  second best among the studied methods.


Defects4J (Version 2.0.0) \cite{just2014defects4j} is a collection of 17 programs and 835 faulty program versions containing a wide spectrum of real software faults.  The number of subject programs and the total number of faulty program versions have nearly doubled in this release of Defects4J. Although the very low failure rates of many Defects4J programs make them non-ideal for statistical fault localization\cite{kuccuk2019impact}, its user friendly scripts and continuous support and evolution makes it an invaluable source for empirical studies.

\begin{figure}
  \fbox{\includegraphics[width=1\columnwidth]{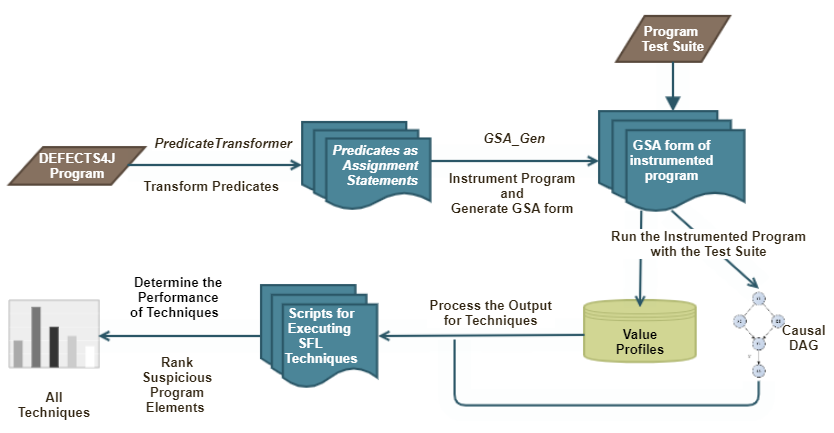}}
  \caption{Process of Empirical Evaluation}
  \label{Pipeline}
\end{figure}

In our study, the techniques we compare assign suspiciousness scores to different program elements. \emph{UniVal} assigns scores to numeric, string, and categorical assignment statements and to predicates, ESP assigns scores to numeric assignment statements and to predicates, and NUMFL assigns scores to each subexpression of numeric assignment statements.
Predicate Switching assigns scores only to predicates. Baah \emph{et al's} linear regression technique \cite{baah2010causal} and the other coverage-based SFL techniques assign scores to all the statements, although each statement within a control dependence region receives the same score. To enable a fair comparison between techniques, we confined the comparison to only predicates and numeric assignment statements.  (Note that to reduce overhead in this study only the faulty classes are instrumented.) With Predicate Switching, however, we reported the cost of finding the nearest predicate to the fault, since the technique does not assign scores to non-predicates.  As a result of these restrictions, our study did not evaluate \emph{UniVal}'s ability to handle string variables.  We intend to do so in a future study.

For a few program versions NUMFL failed to assign a suspiciousness score to any assignment statement and the suspiciousness list only contained zero values.  We suspect this is caused by using a binary outcome variable rather than the absolute difference between the actual and expected output as NUMFL is intended to do.  We did not include these versions in our comparison. We also did not include a program version if it had fewer than 20 ``relevant'' test cases, which cover a faulty class \cite{just2014defects4j}.  The remaining subject program versions are summarized in Table~\ref{tab:subjectprogramsJ}.

\begin{table}

\centering

\begin{tabularx}{\columnwidth}{|l|c|X|l|}

\hline
\multicolumn{4}{|c|}{Defects4J Subject Programs}\\
\hline
Program ID & KLOC & Average \# of tests & \# of faulty versions   \\
\hline

Chart
   & 96           & \multicolumn{1}{|c|}{220}                & 24                  \\ \hline
Cli
 &  4         & \multicolumn{1}{|c|}{147}               & 37                     \\ \hline
Csv
 &  2        & \multicolumn{1}{|c|}{153}               & 16                    \\ \hline
Math
   & 85          & \multicolumn{1}{|c|}{172}                & 103                        \\ \hline
Time
   &   28        &\multicolumn{1}{|c|}{2525}                 & 25                        \\ \hline
Lang
 &  22           & \multicolumn{1}{|c|}{94}               & 60                        \\ \hline
Closure
 &  90          & \multicolumn{1}{|c|}{3448}               & 170                       \\ \hline
Mockito
 &  23         & \multicolumn{1}{|c|}{671}               & 35                      \\ \hline
Codec
 &  10         & \multicolumn{1}{|c|}{132}               & 16                     \\ \hline
JxPath
 &  21         & \multicolumn{1}{|c|}{272}               & 20                     \\ \hline
Gson
 &  12         & \multicolumn{1}{|c|}{566}               & 16                     \\ \hline
 Collections
 &  46        & \multicolumn{1}{|c|}{180}               & 3                    \\ \hline
 Compress
 &  11         & \multicolumn{1}{|c|}{158}               & 45                     \\ \hline

Jsoup
 &  14         & \multicolumn{1}{|c|}{344}               & 90                     \\ \hline
JacksonCore
 &  31         & \multicolumn{1}{|c|}{248}               & 25                   \\ \hline
JacksonXml
 &  6         & \multicolumn{1}{|c|}{143}               & 6                     \\ \hline

JacksonDatabind
 &  4         & \multicolumn{1}{|c|}{1343}               & 109                     \\ \hline

\end{tabularx}
\vspace{2pt}
\caption{Summary of Subject Programs}
\label{tab:subjectprogramsJ}
\vspace{-20pt}
\end{table}

We used the EXAM score measure \cite{wong2008crosstab} and Hit@N measure (sometimes referred to as Recall@N or Top-N)\cite{wang2014version} to report the cost of fault localization for each technique. EXAM score is the percentage of program statements a developer must examine, in non-increasing order of their suspiciousness scores, before finding the fault. If there are ties in the scoring of statements, we assumed that half of the tied statements will have to be examined before a programmer localizes any fault among them.  Hit@N is the number of program versions for which a fault was found within the top N ranked statements in non-increasing order of suspiciousness scores. Because our comparison involves assignment statements and predicates, if the fault was not directly related to an assignment statement or predicate (e.g. a fault of omission) we determined a set of fault localization candidates with an approach similar to the one described by Pearson \emph{et al.} \cite{pearson2017evaluating}. Finally,  We used a linux Ubuntu 20.04 LTS machine that runs on an Intel i5-930H quad core CPU at 2.4-4.1 GHz and that has 8GB of RAM for our experiments. The time measurements are calculated by inserting a timestamp at the beginning and at the end of each step of the pipeline script and reporting the averages for all programs. The EXAM and Hit@N metrics involve simple average or count calculations; hence, we used Microsoft Excel to compute them. 
We ran all the program versions in our comparison 10 times and averaged the results over all runs in case techniques displayed random variation.

\subsection{Results}

Table \ref{tbl:java-table} provides an summary of the results of the empirical evaluation showing the average performance of each technique on each program using the three evaluation methods: EXAM, Hit@10, and Hit@5. Lower EXAM scores are better than high ones, while higher Hit@10 and Hit@5 scores are better than low ones. 
Average runtimes for the methods in our empirical evaluation were: 55.6 seconds for \emph{UniVal}, 78.6 seconds for NUMFL (cumulative time to run both models), 660 seconds for Predicate Switching, 44.8 seconds for Baah2010, 13.2 seconds for ESP,  and 4.4 seconds for the coverage based techniques. The average overhead of our instrumentation is about 25\% (e.g. Math-1 takes 8 seconds to execute without the instrumentation and 10 seconds with the instrumentation).

We make two observations about \emph{UniVal}'s overall performance as shown in Table \ref{tbl:java-table}. First, \emph{UniVal} usually had the best score for a given program/cost metric combination.  Second, as can be seen in the EXAM Score table, \emph{UniVal}'s scores display less variation than other methods. This indicates not only that \emph{UniVal} is capable of relatively precise localization (as shown in the Hit@5 table) but also that it provides more consistent results. In contrast, the well known Ochiai method provides precise localization somewhat frequently, but it exhibits much more variation. 

\begin{table}[t!]
  \includegraphics[width=\columnwidth]{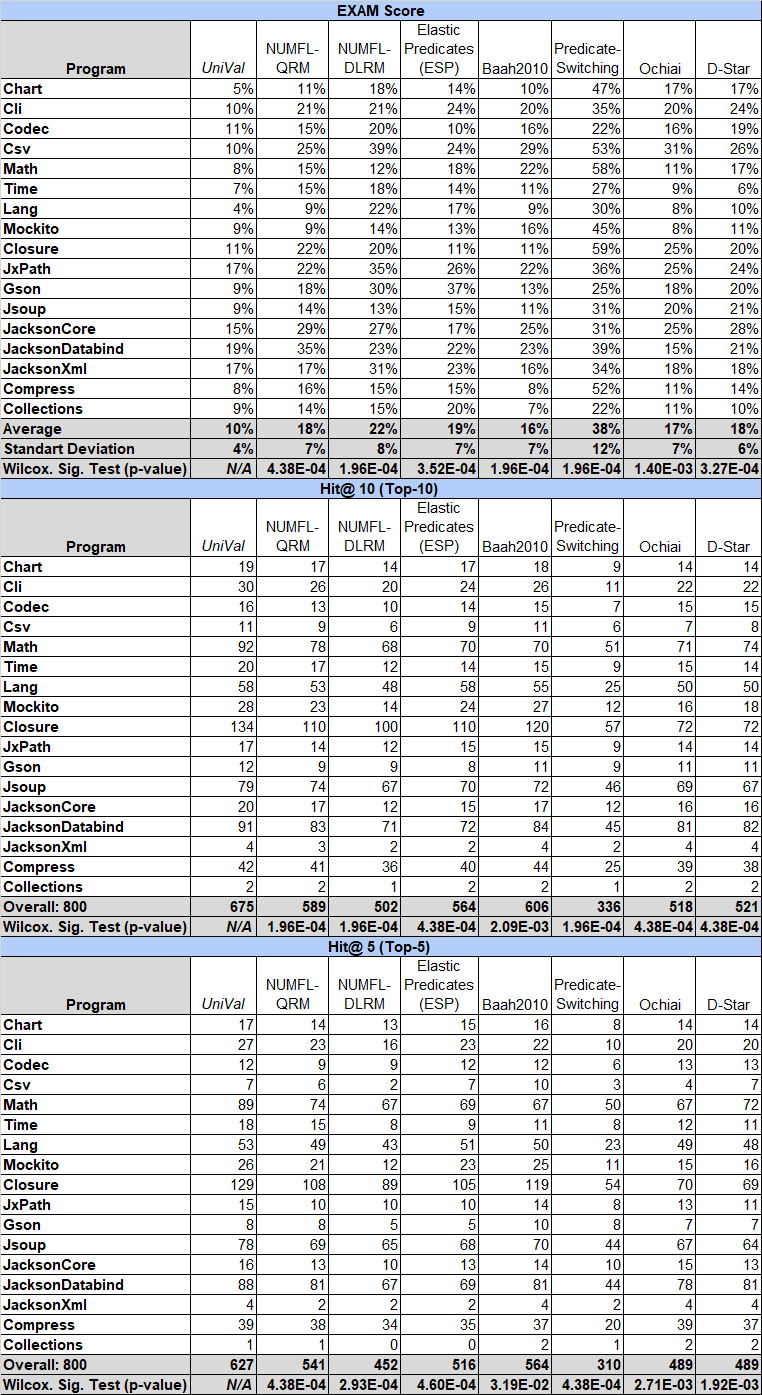}
  \caption{Comparison of \emph{UniVal} and competitive metrics for all Defects4J programs}
  \label{tbl:java-table}
  {\vskip -2em}
\end{table}

To evaluate the statistical significance of our results, we conducted Wilcoxon signed-rank tests \cite{wilcoxon1992individual} for the difference in performance between \emph{UniVal} and other techniques, for each part of Table \ref{tbl:java-table}. The resulting p-values are reported in the bottom row of each part of the table.  Each difference in performance between \emph{UniVal} and another technique is significant at the 0.05 level.

\begin{figure*}[t!]
\centering
  \fbox{\includegraphics[width=\textwidth]{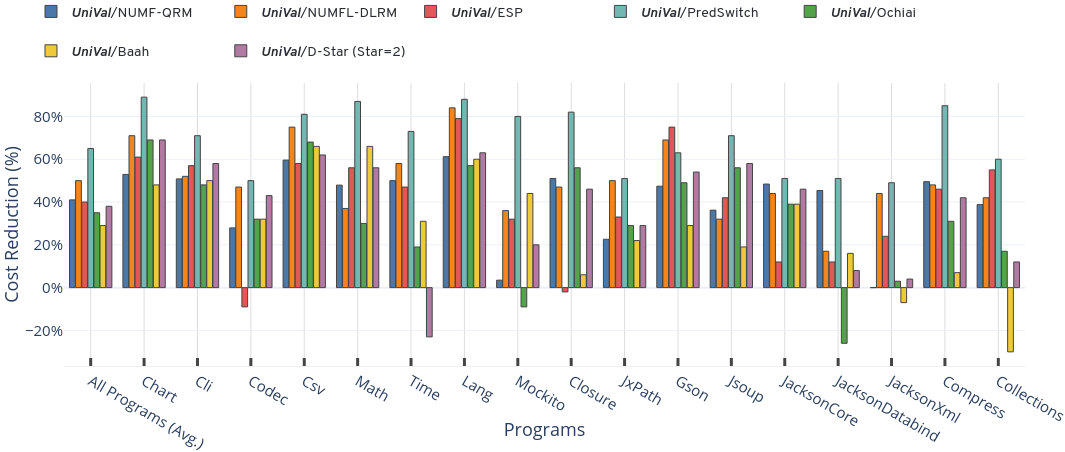}}
 \caption{\emph{EXAM} score reductions achieved by \emph{\emph{UniVal}} over other techniques, for all Defects4J programs}
\label{fig:defects_figure}
\end{figure*}

Figure \ref{fig:defects_figure} visually compares \emph{UniVal}'s performance versus the other techniques using the EXAM score. There are only seven instances in which \emph{UniVal} does worse than a competing technique (out of 119 total).  In general, this study indicates that of all the techniques considered,  \emph{UniVal} has the best performance on the the Defects4J dataset.

\subsection{The Effect of Covariate Balance}

In causal inference, covariates are variables other than the treatment variable or the outcome variable that may be associated with either variable.  They are used for confounding adjustment or for reducing the variance of estimates.  In the absence of confounding, as in a randomized experiment, the joint distributions of the covariates should be similar in both (or in all) treatment groups \cite{hansen2008covariate,hernan2018causal}.  This condition is called \emph{covariate balance}.  It typically does not occur in observational studies.  There are techniques, such as matching \cite{austin2009balance,baah2011mitigating}, that attempt to achieve covariate balance in the analysis sample by removing certain units from the original study sample.  However, \emph{UniVal} statistically adjusts for covariates rather than modifying the study sample.

To better understand the effect of covariate balance and imbalance on the performance of \emph{UniVal} and coverage-based fault localization, we conducted a sub-study in which we measured the degree of covariate imbalance in the data sets for faulty Defects4J\cite{defects4J-dissection} versions and we related it to the cost of fault localization, as measured by EXAM score \cite{wong2008crosstab}, for \emph{UniVal} and the coverage-based SFL metric Ochiai \cite{abreu2009practical}.
To simplify the comparison, we considered only program versions that contain a faulty predicate in a branch condition.  Thus, for both the \emph{UniVal} and Ochiai techniques the treatment value corresponded to the outcome of that predicate (\emph{true} or \emph{false}).  
Additionally, we considered only predicates that have covariates with numerical values. To measure covariate imbalance, we calculated the mean, over all covariates, of the difference in the mean values of individual covariates for the two treatment groups.

We located the program versions in release 1.4 of the Defects4J repository that contained faults in branch predicates by consulting a recent study that details the fault types in that release \cite{defects4J-dissection}. For release 2.0 of Defects4J, we searched among the fault-fix patches included with the release. We found a total of 228 program versions fitting our criteria.

\begin{figure*}[t!]
  \fbox{\includegraphics[width=2\columnwidth]{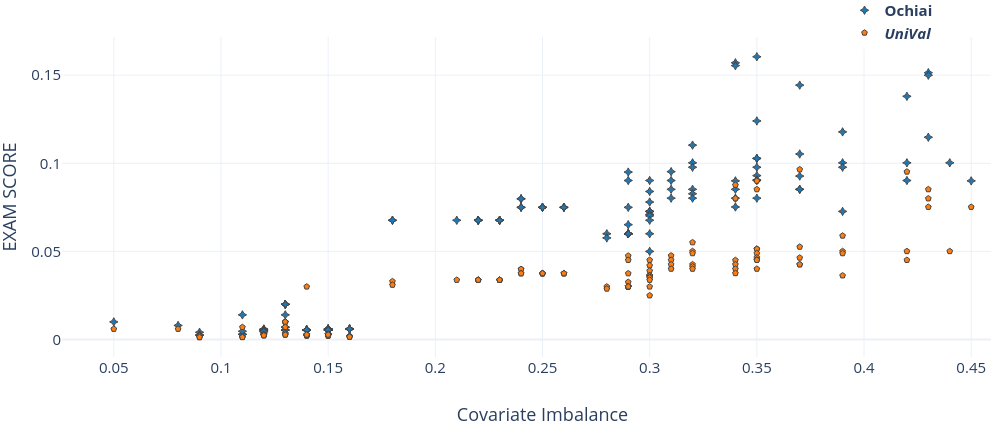}}
  \caption{Relationship Between Covariate Imbalance and EXAM Score}
  \label{CovariateBalance}
  {\vskip -1em}
\end{figure*}

The results of our study of the effect of covariate imbalance on fault localization are depicted in Figure~\ref{CovariateBalance}.  The figure shows a scatter plot in which the X-axis represents the mean, over all covariates, of the difference in the mean values of individual covariates for the two treatment groups and in which the Y-axis represents the EXAM scores for \emph{UniVal} and \emph{Ochiai} for data sets with given levels of covariate imbalance. Note that along the X-axis, larger values represent greater imbalance, and along the Y-axis, larger values represent higher costs for fault localization.

The results indicate that higher fault localization costs are associated with greater imbalance. This is consistent with previous results indicating that covariate balance is associated with lower estimation bias in observational studies \cite{sauppe2017role}. However, it is evident in Figure~\ref{CovariateBalance} that even when the covariates are imbalanced, the cost of fault localization with \emph{UniVal} is usually lower than with Ochiai. This is due to the fact that \emph{UniVal} adjusts for confounding and therefore mitigates the effects of covariate imbalance.

\subsection{Threats to Validity}

\subsubsection{Internal Validity}

As previously noted in the literature the EXAM score and Hit@N metrics are imperfect evaluation methods for automatic fault localization \cite{parnin2011automated,henderson2019evaluating}. In particular they assume a programmer will always start at the top of the ranked list of program elements and move monotonically down the list ignoring the surrounding program structure.  They also assume that the programmer is equally likely to examine program elements with the same suspiciousness score.  But, programmers use their intuition, background knowledge of the program, and other methods of debugging when using automatic fault localization tools. Hence, this evaluation method does not fully account for programmer behavior.

The EXAM and Hit@N cost metrics handle program elements with the same suspiciousness scores by giving each such element the same (average) rank score (see Standard Rank Score in \cite{henderson2019evaluating}).  This creates evaluation bias in favor of fault localization techniques that are more likely than others to assign different elements the same suspiciousness score (such as the coverage-based metrics Ochiai and DStar).  However, some of the techniques we studied --- including the proposed technique \emph{UniVal} --- don't tend give the same scores to different program elements.

\subsubsection{External Validity}

The Defects4J dataset \cite{just2014defects4j} is a very useful collection of programs, bugs, and programmer-written tests. However, no collection can be all encompassing, and this collection is still only a small sample of real programs and their bugs. A technique performing well on these 17 programs might not perform particularly well on any other program (and vice versa). Second, the test cases are provided by the developers and are generally in the form of unit tests.  Arguably, end-to-end system tests would be a better choice for evaluating fault localization techniques  \cite{kuccuk2019impact}.  In particular the tests fail at very low rates on the faulty program versions in the dataset. Ideally, there would be more balance between failing and passing tests.  Finally, because the tests are generally unit tests they do not simulate operational input by end-users the programs.

In general, despite the weaknesses outlined above, the authors feel this study is a state-of-the-art empirical evaluation of fault localization techniques. The study employs only real bugs, real tests, and substantially sized programs. It does not use the often criticized ``injected faults'' or ``generated test cases.'' Nor was it conducted on toy programs constructed for the purpose.  Finally, every effort was made to put the previous work in the best light: from improvements to Baah's method to filtering out faults which NUMFL could not localize.

\section{Related Work}
\label{related_work}
Cleve and Zeller \cite{cleve2005locating} proposed an interventional approach to fault localization based on \emph{cause transitions}, points in time where a variable starts to become the cause of failure, and they showed how delta debugging \cite{zeller2002simplifying} can be used to find them.  Jeffrey \emph{et al.} \cite{jeffrey2008fault} presented a value-based fault localization technique called \emph{value replacement}, which searches for program statements whose execution can be intervened upon to change incorrect program output to correct output.  Similarly, earlier work by Zhang \emph{et al.} introduced \emph{predicate switching}\cite{zhang2006locating}, which searches predicates that are executed by failing tests and whose outcomes can be altered to make the program succeed.  Johnson \emph{et al.} \cite{johnson2020causal} present an approach to ``causal testing'', in which input fuzzing is used to generate passing and failing tests that are similar to an original failing test. The generated tests are then used to pinpoint failures.  The aforementioned techniques each require a complete or partial test oracle.  Furthermore, they also entail possibly costly repeated runs of subject programs and searches among them, while \emph{UniVal} does not require any oracle or repeated runs of subject programs.

Fariha \emph{et al.} \cite{fariha2020causality} propose a technique called Automated Interventional Debugging (AID) that seeks to pinpoint the root cause of an application’s intermittent failures and to generate an explanation of how the root
cause triggers them. AID approximates causal relationships between events in terms of their occurrence times and intervenes
to change the value of predicates in order to prune a causal DAG and extract a causal path from the root cause to a failure.  Although AID makes use of counterfactuals, it does not adjust for confounding bias.  

Baah \emph{et al.}\cite{baah2010causal} pointed out that the conventional SFL techniques are susceptible to confounding bias, and they employed  causal inference methodology to localize faults.  A linear regression model was fitted for test outcomes with a coverage indicator for the target statement as the treatment and a coverage indicator for its forward control dependence predecessor as a covariate.  Bai \emph{et al.} presented two variants of a value-based causal statistical fault localization technique called NUMFL \cite{bai2017causal}. This technique makes use of \emph{generalized propensity scores}, which are used to achieve covariate balance. Although these causal inference based techniques adjust for confounding like \emph{UniVal}, they employ parametric regression and thus lack the modeling flexbility of random forests.  Elastic predicates (ESP) \cite{gore2011statistical}, which were presented by Gore \emph{et al.}, involve measuring how different, in standard deviations, an assigned variable's value is from its average value.  This difference is used as a suspiciousness score.  In contrast to \emph{UniVal}, ESP does not control for confounding bias.

Feyzi \emph{et al.} proposed an approach\cite{feyzi2019inforence} to reducing the number of statements that must be considered in fault localization. They use backward slicing techniques and information theory to find candidate cause-effect chains (introduced by Zeller \emph{et al.}\cite{zeller2002simplifying}), before applying causal inference based techniques (such as Baah2010) on these candidates. Their method is still subjected to the limitations of the techniques it combines, which are mentioned throughout this section. 

Recent studies have investigated combinations of SFL metrics or sources of information for use in fault localization. Xuan \emph{et al.} \cite{xuan2014learning} presented a technique that uses learning-to-rank methods \cite{liu2009learning} to combine multiple SFL metrics. Sohn and Yoo \cite{sohn2017fluccs} combined SFL results with source code metrics from static analysis.  Zou \emph{et al.} \cite{zou2019empirical} found that combining a variety of SFL techniques was beneficial.  Li \emph{et al.} \cite{li2019deepfl} uses deep learning with neural networks to integrate multiple sources of information that vary from SFL metrics to textual similarity measures. Although these approaches to combining different sources of information in fault localization are promising, they do not adjust for confounding or other biases and therefore their results are prone to bias.\cite{liu2020replicability} 
Mutation-based SFL techniques \cite{papadakis2015metallaxis,zhang2013injecting,moon2014ask} measure the suspiciousness of a program statement by how much a mutation  to  it  can  change  the  number  of  program failures  that  occur  on  a  test  set.  These techniques often generate a very large number of mutations, hence they may be very costly to apply.  Unlike \emph{UniVal}, they do not employ established causal inference methodology. 

\emph{Model-based debugging} (MBD) techniques apply model-based diagnosis to software fault localization \cite{mayer2008evaluating,mayer2008prioritising,wotawa2012automated,abreu2009new}. In MBD
a logical model of a program is derived automatically from its source code and a search algorithm finds minimal sets of statements whose faultiness can explain incorrect behavior observed on test cases.  Abreu \emph{et al.}\cite{abreu2009new} presented a new model-based approach, named BARINEL, to diagnosing multiple intermittent faults, which uses a Bayesian method for estimating the probability that a faulty component exhibits correct behavior.  Recent studies \cite{jannach2019fragment,abreu2015using,abreu2012debugging} have applied model-based debugging to spreadsheet programs, which are a special type of numerical program. We believe \emph{UniVal} might be an efficient algorithm for spreadsheet programs.  None of the aforementioned MBD techniques addresses confounding bias.  

\section{Conclusion}
This paper has presented \emph{UniVal}, which is a novel approach to statistical fault localization that is based on statistical causal inference methodology and that integrates value-based and predicate-based fault localization by transforming predicates into assignment statements.  It uses a machine learning model to estimate, with minimal bias, the average causal effect of counterfactual assignments to program variables, without actually changing the program or its executions.  \emph{UniVal} currently handles the values of numeric, boolean, categorical, and string values.  We reported the results of an extensive empirical evaluation of \emph{UniVal}, in which it outperformed a variety of competing techniques.  In future work, we intend to expand the range of program elements and data types to which \emph{UniVal} applies.
\section*{Data Availability}
Our implementation for the prototype tools and experiments presented in this paper is publicly available \cite{yigit_kucuk_2021_4441439}.
\section*{Acknowledgement}
This work was partially supported by NSF award CCF-1525178 to Case Western Reserve University. The authors would also like to thank Zhoufu Bai for providing scripts we used for including NUMFL into our evaluation. 

\bibliographystyle{IEEEtran}
\bibliography{bibliography}

\end{document}